\documentstyle[12pt]{article}
\begin{document}
\parindent 1.3cm
\thispagestyle{empty}   % to suppress the page number on the first page
\vspace*{-3cm}
\noindent
\def\sd{\strut\displaystyle}

\begin{obeylines}
\begin{flushright}

UAB-FT-300/93

\end{flushright}
\end{obeylines}

\vspace{2cm}

\begin{center}
\begin{bf}
\noindent
 THE $\pi^0\to e^+e^-$ AND $\eta\to \mu^+ \mu^-$ DECAYS REVISITED
\end{bf}
  \vspace{1.5cm}\\
Ll. AMETLLER
\vspace{0.1cm}\\
Departament F{\'\i}sica i Enginyeria Nuclear, FIB, UPC,\\
08028 Barcelona, Spain\\
\vspace{0.5cm}
A. BRAMON
\vspace{0.1cm}\\
Grup de F\'\i sica Te\`orica, Universitat Aut\`onoma de Barcelona,\\
08193 Bellaterra (Barcelona), Spain\\
   \vspace{0.5cm}
and\\
E. MASS\'O
\vspace{0.1cm}\\
Grup de F\'\i sica Te\`orica
and Institut de F\'\i sica d'Altes Energies,\\
Universitat Aut\`onoma de Barcelona,\\
08193 Bellaterra (Barcelona), Spain\\

\vspace{2cm}

{\bf ABSTRACT}
\end{center}

The rare $\pi^0 \to e^+e^-$ and $\eta \to \mu^+\mu^-$ decays are
calculated in different schemes, which are seen to be essentially
equivalent to and produce the same results as conventional Vector-Meson
Dominance. We obtain the theoretical predictions
$B(\pi^0 \to e^+e^-) = (6.41 \pm 0.19)\times 10^{-8}$ and
$B(\eta \to \mu^+\mu^-) =\sd (1.14^{+0.07}_{-0.03}) \times 10^{-5}$,
where $B(P\to l^+ l^-)=\Gamma(P\to l^+l^-)/\Gamma(P\to \gamma\gamma)$,
in reasonable agreement with recent experimental data.

\newpage
New experimental data on the rare $\eta \to \mu^+\mu^-$ and $\pi^0 \to e^+e^-$
decays have been obtained very recently by different groups,
and more accurate results are expected in the near future. On the one
hand, the $\eta$ tagging facility at Saturne reports \cite{sat} a branching
ratio
$BR(\eta \to \mu^+\mu^-)=\Gamma(\eta \to \mu^+\mu^-)/\Gamma(\eta \to all)=
(5.6^{+0.6}_{-0.7}\pm 0.5)\times10^{-6},$
to be compared to the old result
\cite{dzh} $BR(\eta \to \mu^+\mu^-)=(6.5\pm2.1)\times10^{-6}$.
Normalizing the new Saturne's measurement to $\eta \to \gamma\gamma$ one gets
\begin{equation}
\label{etaexp}
B(\eta \to \mu^+\mu^-) \equiv {\Gamma(\eta \to \mu^+\mu^-) \over
\Gamma(\eta \to \gamma\gamma)} = (1.4 \pm 0.2) \times 10^{-5} \ ,
\end{equation}
where the branching ratio \cite{pdg} $BR(\eta \to \gamma\gamma) =
0.389\pm0.005$
has been used. On the
other hand, the $\pi^0\to e^+e^-$ branching ratio has been recently measured at
Brookhaven \cite{bnl} and at Fermilab \cite{fnal} with the results
$(6.9 \pm 2.4) \times 10^{-8}$ and $(7.8 \pm 3.1) \times 10^{-8}$,
respectively. Averaging these data and using  \cite{pdg}
$BR(\pi^0 \to \gamma\gamma) = 0.988$ one similarly has
\begin{equation}
\label{pi0exp}
B(\pi^0 \to e^+e^-) \equiv {\Gamma(\pi^0\to e^+e^-) \over
 \Gamma(\pi^0 \to \gamma\gamma)} = (7.3 \pm 1.9) \times 10^{-8} \ .
\end{equation}

The ``reduced'' ratios (1,2) can be expressed in terms of a dimensionless
``reduced'' amplitude $R(P \to l^+l^-) \equiv R$, normalized to the
intermediate $P \to \gamma\gamma$ amplitude, leading to
\begin{equation}
B(P \to l^+l^-) = 2\beta \left( {\alpha \over \pi} {m_l \over m_P}
\right)^2 |R(P \to l^+l^-)|^2,
\end{equation}
where $\beta = \sqrt{1 - 4m_l^2/m_P^2}$. The on-shell $\gamma\gamma$
intermediate state generates the model-independent imaginary part of $R$
\begin{equation}
{\rm Im}\, R(P \to l^+l^-) = {\pi \over 2 \beta} \ln{1-\beta \over 1+\beta}
\ .
\end{equation}
The unitary bound on $B$, $B \geq B^{unit}$, is then obtained by setting
${\rm Re}\, R = 0$ in (3). It takes the values
\begin{eqnarray}
B^{unit} (\eta \to \mu^+\mu^-) & = & 1.11 \times 10^{-5}   \\
B^{unit} (\pi^0 \to e^+e^-) & = & 4.75 \times 10^{-8} \ .
\end{eqnarray}
In units of $B^{unit}$, the Saturne's result (1) and the average (2) are
\begin{eqnarray}
{B(\eta \to \mu^+\mu^-) / B^{unit}} & = & 1.3 \pm 0.2  \\
{B(\pi^0 \to e^+e^-) / B^{unit}} & = & 1.54 \pm 0.40 \ .
\end{eqnarray}
The values on the rhs of eqs.~(7,8), which
correspond to $1 + ({\rm Re} R/{\rm Im} R)^2$, can be used to extract
${\rm Re} R$ from experiment
\begin{eqnarray}
\label{realeta}
{\rm Re}\, R(\eta \to \mu^+\mu^-)  & = &
\pm \left(3.0^{+0.9}_{-1.2} \right) \\
\label{realpi}
{\rm Re}\, R(\pi^0 \to e^+e^-)  & = &
\pm \left(12.9^{+4.0}_{-6.5}\right) \ .
\end{eqnarray}
We will see below that we are able to choose a sign for ${\rm Re} R$ from
theoretical considerations.
 While the imaginary part of $R$ is
finite, model-independent and dominant, the real part of $R$
contains an {\it a priori} divergent
$\gamma\gamma$ loop, depends on the hadronic physics governing the
$P \to \gamma^*\gamma^*$ transition (with off-shell photons) and,
according to eqs. (7,8), amounts only to a fraction of Im$R$.

Calculations of Re$R$ have been performed by many authors in,
essentially, two different contexts. One consists in using
Vector Meson Dominance (VMD) ideas \cite{VMD}, thus
introducing the corresponding VMD form factor to regularize the photon-photon
loop. Hadronic couplings cancel precisely in the ``reduced''
amplitude $R$, which
turns out to depend essentially only on the vector meson mass $M_V$ in the form
factor.
Alternatively, one can rely on (constituent)
quark model ideas \cite{QM} to
regularize the $P \to \gamma^*\gamma^*$ vertex thus obtaining a finite and
reasonable value for Re$R$. A recent paper by Margolis et al.~\cite{Margolis}
confirms the validity (as well as some degree of model-independency) of this
approach. In both contexts,
one obtains rather stable results which are in reasonable agreement with
the above data. The accuracy and reliability of these methods can obviously be
improved when used to compute differences of two Re$R$'s rather than Re$R$'s
themselves, as shown by the authors \cite{PR} a decade ago. The recent and
partly related paper by Savage, Luke and Wise \cite{Wise}, as well as the
publication of new experimental results,
have prompted us to reconsider the situation.

Assuming the dominance of the two photon contribution, the reduced amplitude
$R(q^2) = R(P \to l^+l^-)$ can be written as (see Ref.~\cite{suec}, where
details can be found)
\begin{equation}
R(q^2) = {2i \over \pi^2 q^2}
\int d^4k {q^2 k^2 - (q\cdot k)^2 \over k^2(q-k)^2[(p-k)^2-m_l^2]} F(k^2,k'^2),
\end{equation}
where $q^2=m_P^2,p^2=m_l^2$ and $k'=q-k$, and $F$ is a generic and
model-dependent form factor, with $F(0,0)=1$ for on-shell photons.
The simplest and more transparent way to fix $F$ is by invoking conventional
Vector Meson Dominace (VMD) ideas. This essentially implies neglecting direct
$P\gamma\gamma$ and (single V) $VP\gamma$ vertices, thus assuming the full
dominance of the (double V) chain $P \to VV \to \gamma\gamma.$ The
form factor in this case is
\begin{equation}
\label{FVV}
F=F_{VV}={M_V^2 \over M_V^2-k^2}{M_V^2
\over M_V^2-k'^2}. \end{equation}
Let us present the theoretical predictions
for the real part of the amplitude in this naive and conventional VMD model.
Taking $M_V=M_{\rho,\omega}=0.77\pm0.10$ GeV (this error will be justified
later),
one gets
\begin{equation}
\label{Ree}
{\rm Re}\, R_{\rho,\omega}(\pi^0 \to e^+e^-) = 10.4 \pm 0.6 .
\end{equation}
Quite independently, one also obtains ${\rm Im}\,R(\pi^0 \to e^+e^-) = -17.5 $
and then the ratio
\begin{equation}
\label{Bee}
 B(\pi^0 \to e^+e^-) = (1.35 \pm 0.04) B^{unit}=
(6.41 \pm 0.19)\times 10^{-8},
\end{equation}
in good agreement with the recent data (\ref{pi0exp}) (but in sharp contrast
with the old data \cite{old} available when similar VMD estimates were commonly
performed).
The $\eta \to \mu^+\mu^-$ decay can be similarly analyzed. In the
$SU(3)$-symmetric limit, i.e., using $M_V=M_{\rho,\omega}$ in (\ref{FVV}) and
ignoring  $M_\phi > M_{\rho,\omega}$,
we get
\begin{equation}
\label{Rmmsu3}
{\rm Re}\, R_{\rho,\omega}(\eta \to \mu^+\mu^-)= -1.3^{+0.7}_{-0.5} \ ,
\end{equation}
whereas ${\rm Im}\, R(\eta \to \mu^+\mu^-)=-5.47$. The predicted
ratio is then
\begin{equation}
\label{Bmmsu3}
 B(\eta \to \mu^+\mu^-) = (1.06^{+0.06}_{-0.05}) B^{unit}=
(1.18^{+0.08}_{-0.06})\times 10^{-5} \ ,
\end{equation}
in reasonable agreement with the corresponding experimental value
(\ref{etaexp}).
To calculate $B(\eta \to \mu^+\mu^-)$ in the more realistic case of $SU(3)$
breaking we use the $\eta-\eta'$ mixing angle $\theta_P=-19.5^o$ \cite{mix},
corresponding to an $\eta$ quark content
$\eta = (u \bar u + d \bar d - s \bar s)/ \sqrt 3$,
and introduce $M_\phi > M_{\rho,\omega}$. This easily leads to
\begin{equation}
\label{etares}
\begin{array}{l}
{\rm Re}\, R(\eta \to \mu^+\mu^-)  = \sd {5 \over 4}\,
{\rm Re}\, R_{\rho,\omega}(\eta \to \mu^+\mu^-)- {1 \over 4}\,
{\rm Re}\, R_{\phi}(\eta \to \mu^+\mu^-)=-1.0
^{+0.9}_{-0.6}  \\
\\
B(\eta \to \mu^+\mu^-) = (1.03^{+0.06}_{-0.03})
B^{unit}= (1.14^{+0.07}_{-0.03}) \times 10^{-5}\ ,
\end{array}
\end{equation}
marginally consistent with the data.
For completeness, we also quote the
corresponding results for the $\eta \to e^+e^-$ decay amplitude:
$\quad {\rm Re}\, R(\eta \to e^+e^-)=31.3\pm2.0, \quad
{\rm Im}\, R(\eta \to e^+e^-)=-21.9$\quad and
$\quad B(\eta \to e^+e^-)=(3.04\pm0.26)B^{unit}=(1.37\pm0.12) \times 10^{-8}$.

Two comments about our results are in order. First,
the reasonable agreement we get allows us to solve the sign ambiguity
when extracting the real part of the amplitude from experiment
in (\ref{realeta},\ref{realpi}): we have to choose the positive value for
${\rm Re}\, R(\pi^0 \to e^+e^-)$ and the negative value
for ${\rm Re}\, R(\eta \to \mu^+\mu^-)$. The discarded values are
$4$ and $3$ experimental standard deviations away from the theoretical
results. Second, as was noticed by the authors in Ref.~\cite{PR}, some of
the uncertainties related to hadronic scales or cutoffs disappear
when considering differences of two Re$R$, as
$Re R_{\pi ee}- Re R_{\eta \mu\mu}$.
{}From (13) and (17) we get the numerical result
\begin{equation}
\label{bona}
Re R_{\pi ee}(m_\pi^2) - Re R_{\eta\mu\mu}(m_\eta^2) = +11.4 \pm 0.4,
\end{equation}
where the smallness of the error comes from a large cancellation of the
uncertainties in $M_V$ taking place
because of the difference in the lhs
\footnote{To further appreciate this effect,
we rewrite our older result \cite{PR} (correcting a misprint)
$Re R_{\pi ee}(m_\pi^2) - Re R_{\eta\mu\mu}(m_\eta^2)
\simeq   3\ln(\Lambda_\eta / \Lambda_\pi)
- 3 \ln(m_\mu / m_e)+\ln(m_e m_\mu / m_\pi m_\eta)
\ln(m_e m_\eta / m_\mu m_\pi)
-  r  \ln(m_\mu / \Lambda_\eta)
\ln(1- m_\eta^2 / \Lambda_\eta^2) + \cdots$,
where $\Lambda_\eta, \Lambda_\pi$ are cutoffs needed to regularize the
integrals and the form factor model dependence is effectively parametrized
by the term proportional to $r$. (The dots refer to negligible
contributions.) Invoking SU(3) symmetry $\Lambda_\eta=\Lambda_\pi$ and
allowing $0\le r \le 1$ we obtained \cite{PR}
$Re R_{\pi ee}(m_\pi^2) - Re R_{\eta\mu\mu}(m_\eta^2)\simeq +12 \pm 2$,
compatible with (\ref{bona}) but with larger uncertainties.}.
Eq.(\ref{bona}) is fully compatible with the experimental value
\begin{equation}
\label{expe}
Re R_{\pi ee}(m_\pi^2) - Re R_{\eta\mu\mu}(m_\eta^2) = +16^{+5}_{-7} ,
\end{equation}
which is deduced from (\ref{realeta},\ref{realpi})
when solving the sign ambiguities
according to our analysis, and adding the errors linearly.

We are well aware that the above VMD calculations could be
(and essentially were)
performed many years ago.
One may argue that VMD is certainly a successful
phenomenological scheme but old-fashioned
and lacking of solid theoretical support. For this reason
we shall now consider more modern approaches, where the interactions among
pseudoscalars, vector mesons and photons
are dictated by well-defined and QCD-rooted
lagrangians. In these contexts, the octet containing the
lightest pseudoscalar mesons  plays the role of the set of Goldstone bosons
originated through the spontaneous symmetry breaking of the QCD lagrangian for
vanishing $u$, $d$ and $s$ quark masses.
We shall discuss three types of models:
the first refers to improved and updated versions of the
non-linear $\sigma$-model, such as Chiral Perturbation Theory (ChPT) \cite{GL},
and the other two refer to more recent attempts
to include vector-mesons in these
chiral lagrangians, particularly, the ``massive Yang-Mills approach''
\cite{Meiss}
and the ``hidden symmetry scheme'' \cite{Bando}. We shall argue that
in the context of these three models
the VMD form factor in (\ref{FVV}) and the corresponding predictions
(with the quoted theoretical errors)
in eqs.~(\ref{Ree}-\ref{etares}) are fully justified.

Let us start discussing the ``massive Yang-Mills approach''
proposed mainly by Meissner and extensively discussed in \cite{Meiss}. Much as
in the VMD case, vector mesons are introduced as a nonet of gauge bosons
through
conventional covariant derivatives in ungauged, chiral lagrangians with a
Wess-Zumino (WZ) term.
Axial-vectors can be introduced with the same procedure, but
(appropriate) mass terms for both types of spin-1 mesons have to be
incorporated
unsatisfactorily by hand. When gauging the WZ-term, several possibilities are
open concerning the relative weights of axial {\it vs} vector mesons. The most
attractive one is due to Bardeen and concentrates all the effects of the
anomaly
in the axial-vector sector. This is also the choice favoured by Meissner
in his extensive review \cite{Meiss}, where it is also shown the total
equivalence of this Bardeen version of the WZ-term with conventional VMD for
the case in hand. Accordingly, in the most favoured version of the ``massive
Yang-Mills''  approach the $\pi^0$ and $\eta$ couplings to $\gamma^*\gamma^*$
are considered to proceed through the $F_{VV}$ form factor (\ref{FVV})
and therefore to reproduce
precisely our previous, VMD predictions (\ref{Ree}-\ref{etares}).

 The alternative but related ``hidden symmetry scheme''
by Bando et al.~\cite{Bando} looks more interesting for our present
discussion. Vector-mesons
are introduced as ``dynamical'' gauge bosons of the hidden local $U(3)_V$
symmetry in the $U(3)_L \times U(3)_R/U(3)_V$ non-linear $\sigma$ model.
The corresponding
vector-meson masses, $M_V$, are now automatically generated inside the model,
while quantum or QCD effects are expected to generate ``dynamically'' their
kinetic terms. Photons and weak-bosons can be finally incorporated as external
gauge
fields. One is then lead to a well defined theory describing the strong
interactions of pseudoscalar and vector mesons at low energy, as well as their
electro-weak ones. In the WZ-sector the scheme contains three free
parameters, two of which ($a_2$ and $a_3$ in the notation of Refs.~\cite{new}
and \cite{newbis}
where more details can be found) are relevant for our purposes.
Their role is to fix the relative weight of the direct $P \rightarrow \gamma
\gamma$ amplitude to the (singly or doubly) V-mediated ones, $P \rightarrow V
\gamma \rightarrow \gamma \gamma$ and  $P \rightarrow VV \rightarrow \gamma
\gamma$. According to the general analysis by Bando et al.~\cite{Bando}, the
preferred values for these parameters are \cite{new}
\begin{equation}
\label{bando}
a_2 = 2a_3 = -3/16 \pi^2 \ ,
\end{equation}
which reproduce complete VMD, i.e., they lead to a
cancellation of the $P\gamma\gamma$ and $P V
\gamma$ vertices containing direct couplings of photon(s) to hadrons.

In our
specific context we are forced to fix $a_3$ to the above value
(\ref{bando}) by simply
requiring that the present approach has to lead to a finite result, i.e., by
assuming that vector mesons alone are enough to render convergent the otherwise
divergent two-photon loop.
The most appropriate way to fix the remaining parameter $a_2$ consists in
adjusting the recent data coming from $\gamma\gamma^* \to P$ production and
involving one (essentially) real photon and a virtual one. The
$k^{*2}$-dependence
of the latter requires a VMD-like form factor with  averaged
(see ref.~\cite{CELLO})
mass parameters $\Lambda = 0.75 \pm 0.03, 0.77 \pm 0.04 $ and
$0.81 \pm 0.04$ GeV for $\pi^0, \eta$ and $\eta'$ production, respectively.
These
values are immediately interpretated in Bando's context just fixing $a_2 + 2a_3
= -3/8\pi^2$, thus reproducing the complete VMD result (\ref{bando}),
and identifying the
mass parameter $\Lambda$ with the vector masses $M_V$. The numerical
coincidence
between these masses shows that we can safely use the physical, PDG
values \cite{pdg}
for the $\rho,\omega$ and $\phi$ masses (the latter being responsible
for the slight increase in $\Lambda$ when going from $\pi^0$ to $\eta$ and
$\eta'$), affected by errors smaller than some 10\%.
Alternatively, we can interpretate the
above $\gamma \gamma^* \to P$ results as requiring the use of the physical,
PDG vector masses but allowing for slight variations of the $a_2$ parameter
(again, of some 10\%) around its VMD central value (\ref{bando}).
In this case, our $P \to l^+l^-$
amplitude proceeds mainly through the $F_{VV}$  form factor (\ref{FVV})
but it is then allowed to have
small contaminations of a similar (single) VMD form factor,
$F_{V}={M_V^2 / (M_V^2-k^2)}$.
The latter has been discussed by several authors in Refs.~\cite{VMD}
and \cite{suec} showing that it
leads to just  slightly smaller values of Re$R$.
In any one of these two alternative
interpretations we obtain for Re$R$ our central values
(\ref{Ree},\ref{Rmmsu3},\ref{etares}),
affected with errors which
are roughly one half of the  quoted ones. The present analysis can be confirmed
invoking the complete set of data on radiative vector meson decays,
$V \to P\gamma$
(the most clean and accurate being $\Gamma(\omega \to
\pi^0\gamma) = 720 \pm 50$ KeV \cite {pdg}), as well as the (less conclusive)
data coming from $\pi^0, \eta \to \gamma l^+l^-$ decays. Somewhat
conservatively,
however, we have enlarged our input error bars on $M_V$ for two
main reasons. One is
due to a single (unconfirmed) measurement of the form factor in $\omega \to
\pi^0 \mu^+\mu^-$ leading to a mass parameter $\Lambda = 0.65 \pm 0.03$ GeV,
well below the expected $M_\rho$. The second reason refers to recent
theoretical analyses suggesting values for $a_2$ and $a_3$ somewhat different
from the VMD ones (\ref{bando}) as required by the last mentioned value of
$\Lambda$ or
as preferred by the attractive and
simplifying ``minimal coupling'' principle of Pallante
and Petronzio \cite{PP}. Accordingly, we have adopted
$M_\rho = 0.77 \pm 0.10$ GeV thus generating the error estimates
quoted in our main
results (\ref{Ree}-\ref{etares}).

Let's finally turn to consider our previous results from the point of view
of Chiral Perturbation Theory (ChPT). As it is well known, ChPT is a
successful effective
theory accounting for strong and electroweak interactions of pseudoscalar
mesons at low energy. It's a non-renormalizable theory
containing an infinite set
of counterterms needed to cancel the divergencies appearing when computing
loop corrections. Very recently, Savage et al.~\cite{Wise} have
discussed the $P \to l^+l^-$ decays using preliminary data
from Saturne to fix the required local
counterterms and then predicting the $\pi^0$, $\eta \to e^+e^-$
branching ratios.  An alternative way to proceed consists in assuming that the
relevant, finite part of the ChPT counterterms are
 saturated (dominated) by the contributions of
meson resonances. This resonance saturation hypothesis was already
suggested in the original papers by Gasser and Leutwyler \cite{GL}, was further
discussed in \cite{Eck} and has been fully confirmed by several
authors. Vector mesons usually play the central role, thus realizing VMD in a
modern context which turns out to be particularly successful in the anomalous
sector of the ChPT  lagrangian.  As shown in Refs.\cite{new}
and \cite{newbis}, vector-meson contributions are fully dominant in this
sector, well above other ChPT corrections such as the finite part of the chiral
loops.
In this sense,
our previous VMD results
on $P \to l^+l^-$ decays can also be considered as rather safe calculations
in
the context of ChPT with resonance saturation. To further illustrate this point
we have computed the ``reduced'' amplitude $R$ in terms of
the local counterterms proposed in Ref.~\cite{Wise} within the same
renormalization scheme, obtaining
\begin{equation}
\label{our}
\begin{array}{l}
{\rm Re}\, R(q^2=m_P^2)
= -\sd{\chi_1(\Lambda)+\chi_2(\Lambda)\over 4} - {5\over 2}+ 3
\ln{m_l \over \Lambda} \\
{}~ \\
\phantom{Re R=(q^2=m_P^2) }
\sd + {1\over 4\beta}\ln^2 {1-\beta\over 1 + \beta}
+ {\pi^2\over 12 \beta} - {1\over \beta}
Li_2\big({\beta-1\over \beta+1}\big),
\end{array}
\end{equation}
where $\Lambda$ is the subtraction
point. We have checked that this result agrees with the amplitude $A$ in
\cite{Wise} (which is
related to our $R$ by $A=-\sd {\alpha R / \pi^2}$) with
a minor modification: the term +11 in eq.(2.8) of Ref.~\cite{Wise} is now
found to
be +7. This preserves all the relevant
results in \cite{Wise} except that the new values of the Saturne experiment
should require a counterterm (for  our $\Lambda = M_\rho=0.77$ GeV) given by
\begin{equation}
\label{countrho}
\chi_1(M_\rho) + \chi_2(M_\rho) = \left\{ \begin{array}{r}  -7^{+4}_{-5} \\
                                                            -31^{+5}_{-4}
\end{array} \right. .
\end{equation}
In turn, the less precise $\pi^0\to e^+ e^-$ available experimental data
translate into
\begin{equation}
\label{countrhopi}
\chi_1(M_\rho) + \chi_2(M_\rho) = \left\{ \begin{array}{r}  -22^{+25}_{-16} \\
                                                            +81^{+16}_{-25}
\end{array} \right. .
\end{equation}
The first (second) value in (22,23) corresponds to fixing the sign
ambiguities for Re$R$ according (contrary) to our amplitudes. Notice that
the first values are consistent with the existence of a unique
counterterm, while the second ones are not and,
consequently, have to be
discarded. This further confirms that our amplitudes represent a good
description for the $P\to l^+l^-$ processes.
Thanks to the resonance saturation hypothesis, we can go one step
further and predict the value of the finite part of these counterterms.
This amounts to choose $M_{\rho,\omega}~=~0.77$~GeV
 as both the subtraction point
$\Lambda$ and the mass $M_V$ appearing in our $F_{VV}$ form factor
(\ref{FVV}).  Our previous results for $\eta \to \mu^+ \mu^-$ (15)
and  $\pi^0\to e^+e^-$ (13) can now be presented as leading to
$\chi_1(M_\rho) + \chi_2(M_\rho) \simeq  -14 $ and $-12$, respectively,
close to the experimental values displayed in the first row of
(\ref{countrho},\ref{countrhopi}). Notice that we obtain slightly
different cutoff values for the two processes. This is related to the fact that
our VMD saturation hypothesis does not strictly lead to a constant
counterterm, as explicitly required in (10),
but to a function smoothly depending on $M_V$, $m_P$ and $m_l$.

In conclusion, we have performed a careful calculation of the
$\pi^0 \to e^+e^-$ and $\eta \to \mu^+\mu^-$ decay rates in conventional
Vector Meson Dominance. We have shown that the calculation
is equivalent to
those coming from favoured versions of more
 modern approaches such as the ``massive
Yang-Mills approach'' and ``hidden symmetry schemes''. Similarly,
we have predicted the appropriate value for the finite part of the
corresponding ChPT counterterms under the resonance saturation hypothesis.
Special care has been taken when estimating the theoretical errors,
particularly in the rather precise prediction (\ref{bona}) for the
difference between the real part of the two decay amplitudes.
The other two relevant results of our calculation,
eqs.~(\ref{Bee}) and (\ref{etares}),
are in reasonable agreement with recent data.

\vskip1truecm

{\it Acknowledgements.} Discussions with and comments from M. Gar\c con,
R.S. Kessler, B. Mayer, A. Deshpande and Y. Wah (from the Saturne
Collaboration, BNL and FNAL) are warmfully acknowledged.


\newpage




\begin{thebibliography}{12}

\bibitem{sat} R. S. Kessler et al., Phys. Rev. Lett. 70 (93) 892.
 %
\bibitem{dzh} R.I. ~Dzhelyadin et al., Phys. Lett.   87B \ (1980) ~471.
%
\bibitem{pdg}Particle Data Group, Phys. Rev.   D45 \ (1992) ~1.
%
\bibitem{bnl} A. Deshpande, private communication; A. Deshpande et al.,
preprint
(1993) submitted to Phys. Rev. Lett.  %
\bibitem{fnal}Y. Wah, private comunication, and results presented at the DPF
conference at Fermilab, Nov. 1992.  %
\bibitem{VMD} S.D. ~Drell,  Nuovo Cim.   11 \ (1959) ~471.\\
S.M. ~Berman and D.A.~Geffen, ibid  18 \ (1960) ~1192.\\
B.L.~Young, Phys. Rev.  161 \  (1967) ~1620.\\
C. ~Quigg and J.D. ~Jackson, UCRL Report No. 18487 (1968) (unpublished).\\
A. Ivanov and V. M. Shekhter, Sov. J. Nucl. physics. 32 (1980) 410.\\
L. ~Bergstr\"om, Z. Phys.  C14 \  (1982)~129.\\
L. ~Bergstr\"om et al., Phys. Lett.  126B \ (1983)~117.\\
L. ~Bergstr\"om and E. ~Ma, Phys. Rev.  D29 (1984) ~1029.\\
G.B. ~Tupper and M.A. ~Samuel, Phys. Rev.  D26 \  (1982) ~3302; ibid
29 \ (1984) ~1031.
%
\bibitem{QM} M. ~Pratap and J.~Smith, Phys. Rev.  D5 \ (1972) ~2020.\\
Ll. ~Ametller et al., Nucl. Phys.  B228 (1983) ~301.\\
A. Pich and J. Bernab\'eu, Z. Phys. C22 (1984) 197.

\bibitem{Margolis} M. Margolis et al., TRIUMF TRI-PP-92-99 preprint (1992).

\bibitem{PR} Ll. ~Ametller, A. ~Bramon and E. ~Mass\'o,  Phys. Rev.  D30
(1984) 251.

\bibitem{Wise} M.J.~Savage, M. ~Luke and M.B. ~Wise, Phys. Lett.  291B \
(1992) 48.
%
\bibitem{suec}L. Bergstr\"om et al. in Ref.\cite{VMD}.
%
%G. ~Triantaphyllov, Yale Univ. preprint (1993).
%
\bibitem{old} R.E. ~Mischke et al., Phys. Rev. Lett.  48 (1982) 1153.\\
J. ~Fishcher et al., Phys. Lett.  73B \ (1978) ~364; ibid  76B \  91978)
663 (E).
%
\bibitem{mix} F. J. Gilman and R. Kaufman, Phys. Rev. D36 (1987) 2761\\
A. ~Bramon, Phys. Lett. 51B (1974) 87.
%
\bibitem{GL} J.Gasser and H.Leutwyler, Nucl. Phys. B250 (1985) 465 and 517.
%
\bibitem{Meiss} U-G. Meissner, Phys. Rep. C161 (1988) 213. \\
See also \"O. Kaymakcalan and J. Schechter, Phys. Rev. D31 (1985) 1109.
%
\bibitem{Bando} M. Bando et al. Phys. Rep. C164 (1988) 217 and Phys. Lett.
297B (1992) 151.
%
\bibitem{new} J. Bijnens, A. Bramon and F. Cornet, Z. Phys. C46 (1990) 599.\\
J. Bijnens, A. Bramon and F. Cornet, Phys. Lett. 237B (1990) 488.
%
\bibitem{newbis}
A. Bramon, A. Grau and G. Pancheri, Phys. Lett. B  (1992).\\
A. Bramon, E. Pallante and R. Petronzio, Phys. Lett. 271B (1991) 237.
%
\bibitem{CELLO} CELLO Collaboration, H. J. Behrend et al. Z. Phys. C49
(1991) 401.
%
\bibitem{PP} E. Pallante and R. Petronzio, Phys. Lett. 292B (1992) 143;
Rome Preprint ROM2F 92/37 and Nucl. Phys. (to appear).
%
\bibitem{Eck} G. Ecker et al. Nucl. Phys. B321 (1989) 311. \\
G. Ecker et al. Phys. Lett 223B (1989) 425. \\
J.F. Donoghue, C. Ramirez and G. Valencia, Phys. Rev. D39 (1989) 1947.
%
\end{thebibliography}
\end{document}